\documentstyle[12pt]{article}
\def\~{\tilde}
\def\e{\epsilon}

\def\a{\begin{eqnarray}}
\def\b{\end{eqnarray}}
\def\0{\nonumber}
\def\ba{\begin{array}}
\def\ea{\end{array}}

\renewcommand{\theequation}{\thesection.\arabic{equation}}

\newlength{\extraspace}
\newlength{\extraspaces}
\newcounter{dummy}
\newcommand{\ai}{
\addtocounter{equation}{1}
\setcounter{dummy}{\value{equation}}
\setcounter{equation}{0}
\renewcommand{\theequation}{\thesection.\arabic{dummy}\alph{equation}}
\begin{eqnarray}
\addtolength{\abovedisplayskip}{\extraspaces}
\addtolength{\belowdisplayskip}{\extraspaces}
\addtolength{\abovedisplayshortskip}{\extraspace}
\addtolength{\belowdisplayshortskip}{\extraspace}}
\newcommand{\bj}{
\end{eqnarray}
\setcounter{equation}{\value{dummy}}
\renewcommand{\theequation}{\thesection.\arabic{equation}}}
\def\d{{\partial}}
\begin{document}
\begin{flushright}
SISSA-ISAS 76/96/EP\\
IC/96/78\\
IFT-P.014/96\\
hep-th/yymmxxx
\end{flushright}
\vskip0.5cm
\begin{center}
{\LARGE\bf Toda lattice field theories, discrete $W$ algebras, 
Toda lattice hierarchies and quantum groups}\\[1cm]

{\large L. Bonora$^{1}$, L.P.Colatto$^{2}$ and C.P.Constantinidis$^{3}$}
{}~\\
\quad \\
{\em ~$~^{(1)}$ International School for Advanced Studies (SISSA/ISAS),}\\
{\em Via Beirut 2, 34014 Trieste, Italy, and INFN, Sezione di Trieste}\\
{\em ~$~^{(2)}$ International Centre for Theoretical Physics, P.O. Box 586, 34014 Trieste, Italy }\\
{\em ~$~^{(3)}$ Instituto de F\'\i sica Te\'orica -- UNESP,
 Rua Pamplona 145, 01405 ${\rm S\tilde ao}$ Paulo, Brasil}\\
\end{center}

\vskip 3cm
\abstract{In analogy with the Liouville case we study the $sl_3$ Toda theory 
on the lattice and define the relevant quadratic algebra and out of it we 
recover the discrete $W_3$ algebra. We define an integrable system 
with respect to the latter
and establish the relation with the Toda lattice hierarchy. We compute the
the relevant continuum limits. Finally we find the quantum version of the
quadratic algebra.} 
\vfill\eject

\section{Introduction}

It is well--known that one can define and quantize the
Liouville field theory on a lattice, \cite{FT},\cite{Vol},\cite{B2},\cite{BB}.
Following the last two references, it was shown in \cite{BoBo} that the same
can be done with more general Toda field theories. Basically one
exploits both the integrability properties of such theories, which carry over 
to the lattice in a straightforward way, and the conformal structure 
which reappears on the lattice in a discrete but recognizable way (see
\cite{FG}). As for the quantum theory, one heavily relies on the 
quantum group representation theory.

In this paper we concentrate mostly on the discrete algebras that underlie
the Toda lattice theories and on the relation with integrable hierarchies
and matrix models. It is 
true that one can find analogous algebras and relations
in a continuous parallel treatment of these models, but it is striking to find
a kind of universality in a lattice approach (see for example the coincidence
of our algebra (\ref{disw3}) with that of \cite{BC}).

The paper is organized as follows. In section 2, of pedagogical character, we
collect mostly well--known statements and results about the discrete Liouville 
model  and its connection with the KdV hierarchy and the 1--matrix model.
In section 3, starting from the SL(3) Toda field theory on the lattice,
we compute the discrete $W_3$ algebra and construct an associated 
integrable model; the latter turns out to be a sort of linearization of 
the Toda lattice hierarchy. In section 4 we present the quadratic 
discrete quantum $W_3$ algebra. Section 5 is devoted to a discussion: we
explain in particular in what sense the problems studied here may be connected
to 2D gravity and string theory.

\section{Liouville theory on the lattice and the KdV case.}
\setcounter{equation}{0}
\setcounter{subsection}{0}

In this section we essentially collect known results concerning the 
Liouville theory on the lattice and one--matrix model from refs.\cite{B2} 
and \cite{BMX}, respectively. In \cite{B2} it was shown that to each chiral half
of a Liouville field theory we can associate a lattice field theory,
which (in analogy with the continuum case, \cite{B1}) is characterized by a 
(canonical) quadratic algebra with 
generators $W_n^{(i)}$, $i=1,2$ and $n$ is the lattice variable (discretized
space). The generators are constructed according to
\a
W_n^{(i)}=\sigma_n^1\sigma_{n+i}^2-\sigma_n^2\sigma^1_{n+i}\label{Wliou}
\b
where $\sigma_n^i$ are the discrete analogs of the two distinct solutions 
of the second order differential equation $(\d^2 +u)\sigma=0$, which underlies
the Liouville theory -- $u$ is the chiral energy--momentum tensor. Then,
the quadratic algebra is
\a
&&
\left\{W_n^{(1)}\, ,\, W_m^{(1)}\right\}\,=\,W_n^{(1)}W_m^{(1)}
(\delta_{n,m-1}-\delta_{n,m+1})\0\\
&&
\left\{W_n^{(1)}\, ,\, W_m^{(2)}\right\}\,=\,\ W_n^{(1)}W_m^{(2)}
(-\delta_{n,m+2}+\delta_{n,m+1}-\delta_{n,m}+\delta_{n,m-1})\0\\
&&
\left\{W_n^{(2)}\, ,\, W_m^{(2)}\right\}\,=\, W_n^{(2)}W_m^{(2)}
(\delta_{n,m-2}-2\delta_{n,m-1}+2\delta_{n,m+1}-\delta_{n,m+2})+
\0\\
&&~~~~~~~~~~~~~~~~~~~~~~~~ -4W_{n-1}^{(1)}W_{n+1}^{(1)}\delta_{n, m+1}+4
W_n^{(1)}W_{n+2}^{(1)}\delta_{n, m-1}\label{WnLiou}
\b
But if we introduce the quantity
\a
S_n= 4 \frac{W_{n+1}^{(1)} W_{n-1}^{(1)}}{W_n^{(2)} W_{n-1}^{(2)}}\label{Sn}
\b
the algebra (\ref{WnLiou}) shows two simple factors: the subalgebra spanned by
$W_n^{(1)}$ and
\a
\left\{S_n\,,\,S_m\right\} &=& S_nS_m \left[(4-S_n-S_m)(\delta_{n,m+1}-
\delta_{n,m-1})\right.\0\\
&&\left. +S_{n+1} \delta_{n,m-2} - S_{n-1}\delta_{n,m+2}\right]\label{SnSm}
\b
while
\a
\left\{W_n^{(1)}\, ,\, S_m\right\}\,=0\0\\
\b
$W_n^{(1)}$ carries antichiral degrees of freedom, which decouple from the
theory in the continuum limit and are an artifact of the lattice regularization.
Therefore the significant algebra is (\ref{SnSm}).

The continuum limit is obtained with the following substitutions
\a
&&i\rightarrow x,\qquad j\rightarrow y,\0\\
&&S_i \rightarrow 1 + \e^2 u(x)+...,\qquad\qquad
S_{i+n}\rightarrow 1+ \e^2\left(\sum_{l=0}\frac{(n\e)^l}{l!}
u^{(l)}(x)\right)\label{approx1}\\
&&\delta_{i,j}\rightarrow \e \delta(x,y), \qquad\qquad
\delta_{i+n,j}\rightarrow \e\left(\sum_{l=0}\frac{(n\e)^l}{l!}
\delta^{(l)}(x,y)\right)\0
\b
We find
\a
\lim_{\e \to 0}\frac{1}{4\e^4} RHS = \{u(x),u(y)\}_2\label{kdvlim}
\b
$RHS$ means the right hand side of (\ref{SnSm}) after the substitution, and
\a
\{u(x),u(y)\}_2 = (2u(x)\d_x +u'(x) + {1\over 2}\d_x^3 )\delta(x,y)
\label{KdVPB2}
\b
is the well--known second KdV Poisson structure.

It is remarkable that the algebra (\ref{SnSm}) is met in one--matrix models
with even potential, \cite{BMX}. In this case the semiinfinite matrix $Q$,
which represents the multiplication by the eigenvalue in the orthogonal
polynomial approach, takes the following form:
\a
Q=\sum_{i\geq 0} \Big( E_{i,i+1}+  R_iE_{i,i-1}\Big), 
\qquad\qquad (E_{ij})_{k,l}= \delta_{i,k}\delta_{j,l}\0
\b
One finds two compatible Poisson structures:
\a
\{R_i, R_j\}_1 &=& R_iR_{i+1}\delta_{j,i+1}-
R_iR_{i-1} \delta_{j,i-1}\label{RR1}\\
\{R_i,R_j\}_3 &=& R_iR_{i+1} (R_i+R_{i+1})\delta_{j,i+1} - R_iR_{i-1}
(R_i+R_{i-1}) \delta_{j,i-1} \label{RR2}\\
&&R_iR_{i+1}R_{i+2}\delta_{j,i+2}-R_iR_{i-1}R_{i-2}\delta_{j,i-2}\0
\b
These are the two homogeneous components of the algebra (\ref{SnSm}).
More precisely we have
\a
\{\cdot, \cdot\} = \{\cdot, \cdot\}_3 - 4\{\cdot,\cdot\}_1\0
\b
 
One may wonder whether it is possible to define an infinite set of conserved
quantities and integrable flow equations in correspondence with the 
algebra of the
$S_n$'s or $R_n$'s, which we formally identify in the following.
The answer is easily provided by the matrix model. The hamiltonians are
defined as $H_{2k}= tr Q^{2k}$, so that
\a
H_2 = \sum_n R_n, \qquad H_4 = \sum_n ( {1\over 2} R_n^2 + R_n R_{n+1}),...\0
\b 
and the flows are given by
\a
\frac{\d Q}{\d \tau_k} = [Q^k_+,Q]_-\label{todaKdVflows}
\b
where the -- subscript means that we are considering only the strictly lower 
triangular part of the semiinfinite matrix, while + denotes the upper
triangular part (including the main diagonal). In particular
\a
\frac {\d R_n}{\d t_2} &=&  R_n(R_{n+1}-R_{n-1})\equiv \{R_n , H_2\}_1 \0\\
\frac {\d R_n}{\d t_4} &=&  R_nR_{n+1}\Big(R_n+R_{n+1}+R_{n+2}\Big) - 
 R_nR_{n-1}\Big(R_n+R_{n-1}+R_{n-2}\Big)\equiv\0\\
 &&\{R_n , H_4\}_1 = \{R_n, H_2\}_3 \label{diskdvfl}
\b
In the continuum limit we find
\a
&&\lim_{\e\to 0}\Big(\frac{1}{2\e^3} \frac {\d R_n}{\d t_2}\Big) = u' \equiv
\frac{\d u}{\d \tilde t_1}\0\\
&&\lim_{\e\to 0}\Big[\frac{2}{\e^5}\Big(-\frac {\d R_n}{\d t_2}+ \frac{1}{4}
\frac {\d R_n}{\d t_4}\Big)\Big] 
=u''' + 6uu' \equiv \frac{\d u}{\d \tilde t_3}\0
\b
and so on. The RHS's define the renormalized couplings $\tilde t_1, \tilde t_3,
...$.

Therefore we can call eqs.(\ref{diskdvfl}) discrete KdV flows.

\section{Discrete $W_3$ and discrete Boussinesq}
\setcounter{equation}{0}
\setcounter{subsection}{0}

In ref.\cite{BoBo} it was shown that discretizing the SL(3) Toda theory
on the lattice produces (canonically) a quadratic algebra with generators
$W_n^{(i)}$, $i=1,2,3$, defined as follows
\a
W_n^{(i)}= (-1)^i\varepsilon^{ijkl}\sigma^1_{n+j}\sigma^2_{n+k}\sigma^3_{n+l},
\qquad i,j,k,l=0,1,2,3\label{Wbouss}
\b
where $\varepsilon^{ijkl}$ is the standard constant completely 
antisymmetric tensor; it
follows, in particular, that $W_n^{(0)}\,=\,W_{n+1}^{(3)}$. The $\sigma_n^i$ 
are the discrete analogs of the three distinct solutions of the equation
\a
(\d^3 +u \d +w)\sigma=0\label{cubic}
\b
where $u,w$ are constructed out of the Toda fields. The quadratic algebra is
\a
&&
\left\{W_n^{(1)}\, ,\, W_m^{(1)}\right\}\,=\,-\frac{1}{3}W_n^{(1)}W_m^{(1)}
(\delta_{n,m-3}-\delta_{n,m-1}+\delta_{n,m+1}-\delta_{n,m+3})+
\0\\
&&~~~~~~~~~~~~~~~~~~~~~~~~ +W_n^{(2)}W_{n+2}^{(3)}\delta_{n, m-1}-
W_{n+1}^{(3)}W_{n-1}^{(2)}\delta_{n, m+1}
\0\\
&&
\left\{W_n^{(1)}\, ,\, W_m^{(2)}\right\}\,=\,-\frac{1}{3}W_n^{(1)}W_m^{(2)}
(\delta_{n,m-3}+\delta_{n,m-2}-2\delta_{n,m-1}+
\0\\
&&~~~~~~~~~~~~~~~~~~~~~~~~ +\delta_{n,m}-\delta_{n,m+1}+
\delta_{n,m+2}-\delta_{n,m+3})+ 
 \label{Wn}\\
&&~~~~~~~~~~~~~~~~~~~~~~~~ +W_n^{(3)}W_{n+2}^{(3)}\delta_{n, m-1}-
W_{n+1}^{(3)}W_{n-2}^{(3)}\delta_{n, m+2}
\0\\
&&
\left\{W_n^{(1)}\, ,\, W_m^{(3)}\right\}\,=\,-\frac{1}{3}W_n^{(1)}W_m^{(3)}
(\delta_{n,m-3}+\delta_{n,m-2}-\delta_{n,m-1}-\delta_{n,m+2})
\0\\
&&
\left\{W_n^{(2)}\, ,\, W_m^{(2)}\right\}\,=\,-\frac{1}{3}W_n^{(2)}W_m^{(2)}
(\delta_{n,m-3}-\delta_{n,m-1}+\delta_{n,m+1}-\delta_{n,m+3})+
\0\\
&&~~~~~~~~~~~~~~~~~~~~~~~~ -W_n^{(3)}W_{n+1}^{(1)}\delta_{n, m-1}+
W_n^{(1)}W_{n-1}^{(3)}\delta_{n, m+1}
 \0\\
&& 
\left\{W_n^{(2)}\, ,\, W_m^{(3)}\right\}\,=\,-\frac{1}{3}W_n^{(2)}W_m^{(3)}
(\delta_{n,m-3}+\delta_{n,m}-\delta_{n,m+1}-\delta_{n,m+2})
 \0\\ 
&& 
\left\{W_n^{(3)}\, ,\, W_m^{(3)}\right\}\,=\,-\frac{1}{3}W_n^{(3)}W_m^{(3)}
(\delta_{n,m-2}+\delta_{n,m-1}-\delta_{n,m+1}-\delta_{n,m+2})\0
\b

As in the previous section this algebra contains two irreducible factors.
One is the subalgebra generated by $W_n^{(3)}$, the other is generated
by $S_n$ and $W_n$, which are defined as follows:
\a
S_n = - \frac{W_{n-1}^{(2)}W_{n+1}^{(3)}}{W_{n-1}^{(1)}W_n^{(1)}},\qquad
W_n = Z_nS_nS_{n+1},\qquad 
Z_n= \frac{W_{n}^{(1)}W_{n-1}^{(3)}}{W_{n-1}^{(2)}W_n^{(2)}}
\label{SnWn}
\b
It is lengthy but straightforward to prove that \footnote{This verification
has been first done by V.Bonservizi.}  
\a
\{W_n^{(3)}, S_m\}=0,\qquad\qquad\{W_n^{(3)}, W_m\}=0\0
\b
and
\a
&&\{S_n\, ,\,S_{n+1}\} \,= \,\left(S_n S_{n+1}-W_n\right)\left(1-S_n-S_{n+1}
\right)\0\\
&&\{S_n\, ,\,S_{n+2}\}\,=\, -S_nS_{n+1}S_{n+2} + W_nS_{n+2} +W_{n+1}S_n\0
\b
\a
&&\{W_n\, , \,S_{n-1}\} \,=\, -W_nS_{n-1}\left(1-S_n-S_{n-1}\right)
-W_nW_{n-1}\0\\
&&\{W_n\, , \,S_{n-2}\} \,=\, W_nS_{n-1}S_{n-2}-W_nW_{n-2}\0\\
&&\{W_n\, ,\, S_n\} \,=\, -\{W_n\, ,\, S_{n+1}\} \,=\, 
W_nS_nS_{n+1} -W_n^2\label{disw3}\\
&& \{W_n\, ,\, S_{n+2}\}\, =\, W_nS_{n+2} \left(1-S_{n+1}-S_{n+2}\right) 
+ W_nW_{n+1} \0\\
&&\{W_n\, ,\, S_{n+3}\}\, =\, -W_nS_{n+2}S_{n+3}+ W_nW_{n+2} \0
\b
\a
&&\{W_n\, ,\,W_{n+1}\}\,=\, W_nW_{n+1}\left(1-S_n-S_{n+2}\right)\0\\
&&\{W_n\, ,\,W_{n+2}\}\,=\, W_nW_{n+2}\left(1-S_{n+1}-S_{n+2}\right)\0\\
&&\{W_n\, ,\,W_{n+3}\}\,=\, -W_nW_{n+3}S_{n+2}\0
\b
The other brackets are either obtained in an obvious way from these
or else vanish.

Motivated by what follows we call this the discrete $W_3$ algebra. 
It is striking that this algebra was found some time ago by
A.A.Belov and K.D.Chaltikian \cite{BC}, with a different approach based on
Feigin's construction of lattice screening charges. It is common wisdom
that there may by many ways to discretize a continuous system. This example
seems to imply however that we can find universal structures also in
lattice theories. This is due to the fact that conformal and W structures
persist in a discrete form in lattice theories.

\subsection{The continuum limit}

The continuum limit of the discrete $W_3$ algebra is expected to lead 
to the well-known classical continuous $W_3$ algebra. The technical aspects
of such limit do not seem to have been carefully carried out up to now
and are not as straightforward as in the previous section. 
Therefore we spend a few words on them.

We start from the expansions
\a
S_n \rightarrow {1\over 3} + {1\over 9} \e^2 u(x)+ \ldots,\qquad
W_n \rightarrow  {1\over {27}}\Big(1+\e^2 u(x) + \e^3 w(x)\Big)+\ldots
\label{approx2}
\b
They are expected to lead to the following classical $W_3$ algebra
\ai
&&\{u(x), u(y)\}=(2u(x)\d_x+ u'(x)+2\d^3_x)\delta(x,y)\label{w31}\\
&&\{u(x), w(y)\} =(3w(x) \d_x +2w'(x) - \d_x^2 u(x) - \d_x^4)
\delta(x,y)\label{w32}\\ 
&&\{w(x),w(y)\} = \Big(2w'(x)\d_x + w''(x) -
{2\over 3}(u(x)+\d^2_x)(\d_x u(x)+\d^3_x)\Big)\delta(x,y) \label{w33} 
\bj
Then we write the algebra (\ref{disw3}) in terms of Kronecker $\delta$ symbols
(like in (\ref{SnSm})) and replace the RHS's with the relative 
continuous expressions. We denote by $CL(X,Y)$ the RHS of $\{X_n\, ,\, Y_m\}$
after this substitution, where $X_n$ and $Y_n$ stand for either $S_n$ or $W_n$.
We expect to retrieve the continuous algebra by finding suitable combinations
of the $CL(X,Y)$ so that when $\e\to 0$ one gets one of the commutators 
(\ref{w31}--\ref{w33}).

In fact we find
\a
&&\lim_{\e\to 0}\frac{3^4}{\e^4}\Big( - \, CL(S,S)\Big)=\{u(x),u(y)\}\0\\
&&\lim_{\e\to 0} \frac{3^5}{\e^5}\Big(\,CL(W,S)- 3\,CL(W,W)\Big) =
\{u(x),w(y)\}\0\\
&&\lim_{\e\to 0} \frac{3^4}{\e^6}\Big(- CL(S,S) + 3 CL(S,W) + 3 CL(W,S) 
- 9 CL(W,W)\Big) =\{ w(x), w(y)\}\0
\b
It is evident that one can get the same result directly from (\ref{Wn}).

\subsection{An integrable system}

As in the previous section we may wonder whether we can associate an integrable
system to the Poisson structures (\ref{disw3}). We use the plural because, in 
fact, (\ref{disw3}) contains four homogeneous algebras: if we assign to $S_n$ and
$W_n$ degree 2 and 3, respectively, we see that the right hand sides of 
(\ref{disw3}) contains terms of degree 3,4,5,6. Accordingly we have four
Poisson brackets, which we explicitly write down for convenience
\a
\{S_n\,,\,S_{n+1}\}_0 \,=\,-W_n \label{disw30}
\b
\a
&&\{S_n\,,\,S_{n+1}\}_1 \,=\, S_nS_{n+1},\0\\
&&\{W_n\,,\, S_{n-1}\}_1 \,=\, -W_nS_{n-1},\qquad
\{W_n\,,\, S_{n+2}\}_1 \,=\, W_nS_{n+2}\label{disw31}\\
&&\{W_n\,,\, W_{n+1}\}_1 \,=\, W_nW_{n+1},\qquad
\{W_n\,,\, W_{n+2}\}_1 \,=\, W_nW_{n+2}\0
\b
\a
&&\{S_n\,,\,S_{n+1}\}_2 \,=\,W_n(S_n+S_{n+1}),\qquad
\{S_n\,,\,S_{n+2}\}_2 \,=\,W_nS_{n+2}+W_{n+1}S_n\0\\
&&\{W_n\,,\, S_{n-2}\}_2\,=\,-W_nW_{n-2},\qquad
\{W_n\,,\, S_{n-1}\}_2 \,=\, -W_nW_{n-1}\label{disw32}\\
&&\{W_n\,,\, S_{n}\}_2 \,=\,-W_n^2,\qquad\{W_n\,,\, S_{n+1}\}_2\,=\,W_n^2\0\\
&&\{W_n\,,\, S_{n+2}\}_2\,=\,W_nW_{n+1},\qquad 
\{W_n\,,\, S_{n+3}\}_2 \,=\, W_nW_{n+2}\0
\b
\a
&&\{S_n\,,\,S_{n+1}\}_3\,=\,-S_nS_{n+1}(S_n+S_{n+1}),\qquad
\{S_n\,,\,S_{n+2}\}_3 \,=\,-S_nS_{n+1}S_{n+2}\0\\
&&\{W_n\,,\, S_{n-2}\}_3\,=\,W_nS_{n-1}S_{n-2},\qquad
\{W_n\,,\, S_{n-1}\}_3 \,=\, W_nS_{n-1}(S_n+S_{n-1})\label{disw33}\\
&&\{W_n\,,\, S_{n}\}_3 \,=\,W_nS_nS_{n+1},
\qquad\{W_n\,,\, S_{n+1}\}_3\,=\,-W_nS_nS_{n+1}\0\\
&&\{W_n\,,\, S_{n+2}\}_3\,=\,-W_nS_{n+2}(S_{n+1}+S_{n+2}),\qquad 
\{W_n\,,\, S_{n+3}\}_3 \,=\,- W_nS_{n+2}S_{n+3}\0\\
&&\{W_n\,,\, W_{n+1}\}_3\,=\,-W_nW_{n+1}(S_n+S_{n+2}),\0\\
&&\{W_n\,,\, W_{n+2}\}_3\,=\,-W_nW_{n+2}(S_{n+1}+S_{n+2}),\qquad
\{W_n\,,\, W_{n+3}\}_3\,=\,-W_nW_{n+3}S_{n+2}\0
\b
All the other brackets can either be obtained from the above in an obvious way 
or else vanish.

One can define new (inhomogeneous) Poisson structures as follows
\a
\{\cdot,\cdot\}^{(1)}=\{\cdot,\cdot\}_0+ \{\cdot,\cdot\}_1,\qquad
\{\cdot,\cdot\}^{(2)}=\{\cdot,\cdot\}_2+ \{\cdot,\cdot\}_3\0
\b
With respect to the new Poisson brackets it is possible to define
a bi--hamiltonian system \cite{BC}. The first three hamiltonians were 
introduced in \cite{BC}:
\a
&&{\cal H}_1= \sum_n S_n\0\\
&&{\cal H}_2 = \sum_n\Big(W_n -{1\over 2} S_n^2- S_nS_{n+1}\Big)\label{BCham}\\
&&{\cal H}_3 = \sum_n\Big({1\over 3}S_n^3 +S_nS_{n+1}(S_n+S_{n+1}+S_{n+2})
-W_n(S_{n-1}+S_n+S_{n+1}+S_{n+2})\Big)\0
\b
We will shortly see how to generalize these first few cases with a very simple
construction. The two Poisson brackets defined above are compatible with
respect to these hamiltonians {\it provided we disregard all the third
power terms in $W_n$}. 
Then, the flows are defined by
\a
\frac{\d f_n}{\d t_i} \,=\, \{f_n\,, \, {\cal H}_i\}^{(2)}= \{f_n\,,\,
{\cal H}_{i+1}\}^{(1)}\label{BCflows}
\b
where $f_n$ is either $W_n$ or $S_n$. They will contain terms at most
quadratic in $W_n$ due to the above (essential) truncation. We call the model 
characterized by these truncated flows the BC model, \cite{BC}. 
From these discrete flows one can extract the continuous Boussinesq flows 
with the procedure of the previous subsection.
\a
&&\lim_{\e\to 0} \frac{3^3}{\e^3} \,\frac{\d S_n}{\d t_2} = u'(x)
\equiv\frac{\d u}{\d \tilde t_1},\qquad
\lim_{\e\to 0} \frac{3^3}{\e^4} \Big(3 \,\frac{\d W_n}{\d t_2}-
\frac{\d S_n}{\d t_2}\Big) = w'(x)\equiv\frac{\d w}{\d \tilde t_1}\0
\b
and
\a
&&\lim_{\e\to 0} \frac{3^3}{\e^4} \Big(4 \frac{\d S_n}{\d t_2}-
3 \frac{\d S_n}{\d t_3}\Big) = 2w'(x)- u''(x)
\equiv\frac{\d u}{\d \tilde t_2}\0\\
&&\lim_{\e\to 0} \frac{3^3}{\e^5} \Big(-4 \frac{\d S_n}{\d t_2}+
12\frac{\d W_n}{\d t_2}+3\frac{\d S_n}{\d t_3}-9\frac{\d W_n}{\d t_3}\Big) 
= w''(x)- \frac{2}{3} (u(x)u'(x)+ u'''(x)) \equiv\frac{\d w}{\d \tilde t_2}\0
\b
and so on. The rightmost derivatives define the continuous flow parameters.

\subsection{Connection with the Toda lattice hierarchy}

In this subsection we intend to study the relationship between the truncated
BC model and the Toda lattice hierarchy. We shall see that the BC model can be 
reconstructed from the Toda lattice hierarchy provided we consider a 
suitable linearization of the latter. This will automatically establish a 
relation with two--matrix models, see \cite{BX}.

The connection between Toda lattice hierarchy and differential N--KdV 
hierarchies was studied in \cite{BX}. By Toda lattice hierarchy we 
simply mean here the integrable flows
\a
\frac{\d Q}{\d \tau_k} = [Q^k_+,Q]\label{todaflows}
\b
where $Q$ is a semi--infinite matrix.

In particular, as far as the 3--KdV
or Boussinesq hierarchy is concerned, it has been shown that the relevant 
$Q$ matrix has the form
\a
\bar Q=\sum_{j=0}^\infty \Big( E_{j,j+1} + a_jE_{j+l,j}+b_j E_{j+2,j}\Big), 
\qquad (E_{j,m})_{k,l}= \delta_{j,k}\delta_{m,l}\label{jacobi}
\b
For later convenience we
represent the matrix $\bar Q$ by the following operator
\a
\bar Q(j) = e^{\d_0} + a_j e^{-\d_0}+ b_j e^{-2\d_0}\label{Qj}
\b
where $e^{\d_0}$ is the finite shift operator defined by 
$(e^{\d_0}f)_j=f_{j+1}$, for any discrete function $f_j$.

The contact between (\ref{Qj}) and (\ref{jacobi}) is made by acting with the 
former on a discrete function $\xi(j)$; then $\bar Q(j)\xi(j)$ 
is the same as 
the $j$--th component of $\bar Q \xi$, where $\xi$ is a column vector with 
components $\xi(0),\xi(1),...$. We will generally understand the dependence 
on $j$ in (\ref{Qj}).

If we now define
\a
{\rm H}_k = {\rm Tr} (\bar Q^k)= \sum_j\bar Q^k_{(0)}(j)\0
\b
where the subscript $_{(p)}$ denotes the coefficient of $e^{-p\d_0}$ in 
$\bar Q^k(j)$,
and make the identifications
\a
 a_n\,\equiv\,S_n,\qquad\qquad b_n \,\equiv \,W_{n-1}\label{ident}
\b
we find a remarkable recursion relation
\a
\{f_n\,,\,{\rm H}_k\}_0 \,-\,\{f_n\,,\,{\rm H}_{k-1}\}_1\,
+\,\{f_n\,,\,{\rm H}_{k-2}\}_2 -
\{f_n\,,\,{\rm H}_{k-3}\}_3=0\label{recur}
\b
which holds for any $k$, provided we set ${\rm H}_k=0, \,k\leq 1$.

However the matrix $\bar Q$ cannot be just plugged in eq.(\ref{todaflows}), 
because this would lead to inconsistencies. In fact, in related treatments,
Eq.(\ref{jacobi}) is only a starting point, but,
in order to get the 3--KdV flows, one has either to go through a process
of hamiltonian reduction or to introduce suitable corrections to (\ref{jacobi}),
as in \cite{BCV}. Since we want to find a relation between the BC model 
and the Toda lattice hierarchy, it is reasonable to start from (\ref{jacobi})
and (\ref{todaflows}), but one expects to have to introduce suitable 
modifications.

Let us start from the {\it linearized} hamiltonians
\a
H_k = {\rm Tr} (\bar Q^k_\ell)\0
\b
where the subscript $\ell$ means that in $\bar Q^k$ we keep only terms which are
independent of or linear in the field $b$. Then, if  
we find that the BC hamiltonians are reproduced by
\a
{\cal H}_k = (-1)^{k+1}(H_{2k}-H_{2k-1}),\qquad k=1,2,...\qquad H_1=0\label{BChamgen}
\b
This formula, with the identification (\ref{ident}), provides a 
generalization of (\ref{BCham}).

The derivation of the BC flows is more elaborate. First of all let us introduce
the operators $A(j)= a_j e^{-\d_0}$ and $B(j)=b_je^{-\d_0}$. Then we write 
down the linearized flows
\a
\frac{\d \bar Q}{\d \tau_k} =[ (\bar Q^k_\ell)_+, \bar Q]_- +
[\Big(A\bar Q_\ell^{k-1}\Big)_{(0)},B]e^{-\d_0}\label{Todacorr}
\b
 
Next we define 
\a
\frac{\d a_n}{\d \tau_k} \equiv
\Big(\frac{\d a_n}{\d \tau_k}\Big)_a
+\Big(\frac{\d a_n}{\d \tau_k}\Big)_b\label{anflows}
\b
i.e. we split the $a$ flows into a $b$--independent and a $b$--dependent
part. As a consequence of the above definitions it turns out that the 
$b$--dependent part of the $a$ flows vanish for odd $k$. The  
rightmost correction term in (\ref{Todacorr}) affects only the $b$ flows.

Finally, after the substitution (\ref{ident}), we find that the BC flows
can be expressed in terms of the corrected Toda flows (\ref{Todacorr})
as follows
\a
&&\frac{\d S_n}{\d t_k} = \Big(\frac{\d S_n}{\d \tau_{2k}}\Big)_a - 
\frac{\d S_n}{\d \tau_{2k-1}} + \Big(\frac{\d S_n}{\d\tau_{2k-2}}\Big)_b\0\\
&&\frac {\d W_n}{\d t_k} = \frac{\d W_n}{\d \tau_{2n}} - 
\frac{\d W_n}{\d \tau_{2k-1}}\label{BCtodaf}
\b

\section{Quantum discrete $W$ algebra}

Using the quantum exchange relations of ref.\cite{BoBo}, we compute the
quantum version of (\ref{disw3}). First, inspired by the definition of
quantum determinant in ${\rm SL_q(3)}$, we define the quantum $W_n^{(i)}$ 
generators as follows
\a
W_n^{(i)}= (-1)^i\tilde\varepsilon^{ijkl}\sigma^1_{n+j}\sigma^2_{n+k}
\sigma^3_{n+l},\qquad\qquad i,j,k,l=0,1,2,3\label{qWbouss}
\b
where the ordering of the $\sigma_n^i$ is now crucial and 
\a
\tilde\varepsilon^{ijkl} = \varepsilon^{ijkl} q^p, \qquad\quad 
q=e^{-i\hbar}\0
\b
where $p$ is the number of contiguous exchanges to pass from the ordering
$jkl$ to the natural ordering of the same integers.
Then we compute
\a
&&W^{(1)}_n\, \sigma_m \,=\,\sigma_m \,W^{(1)}_n \,q^{-\frac{2}{3}(\delta_{m,n}-
\delta_{m,n+1}-\delta_{m,n+3})}+ A\, \sigma_n\,W_m^{(3)}\delta_{m,n+1}\0\\
&&W^{(2)}_n\, \sigma_m \,=\,\sigma_m\, W^{(2)}_n\, 
q^{-\frac{2}{3}(\delta_{m,n+2}+
\delta_{m,n}-\delta_{m,n+3})}- B \,\sigma_{n+3}\,W_n^{(3)}\delta_{m,n+2}
\label{Wsigma}\\
&&W^{(3)}_n \,\sigma_m \,=\,\sigma_m \,W^{(3)}_n\, q^{\frac{2}{3}(\delta_{m,n+2}-
\delta_{m,n})}\0
\b
where
\a
A= q^{-{2\over 3}} - q^{4\over 3},\qquad B=q^{{2\over 3}} - q^{-{4\over 3}}\0
\b
The exchange relations (\ref{Wsigma}) lead to
\a
W_n^{(1)}\, W^{(1)}_m &=& W^{(1)}_m\, W^{(1)}_n\, q^{\frac{2}{3}(\delta_{m,n+3}-
\delta_{m,n+1}+\delta_{m,n-1}-\delta_{m,n-3})} \0\\
&&-\,A\,W^{(2)}_{n-1}\,W^{(3)}_{n+1}\delta_{m,n-1} 
-\,B\,W^{(2)}_n\,W^{(3)}_{n+2}\delta_{m,n+1}\0\\
W_n^{(1)}\, W^{(2)}_m &=& W^{(2)}_m\, W^{(1)}_n\, q^{\frac{2}{3}(\delta_{m,n+3}+
\delta_{m,n+2}-2\delta_{m,n+1}+\delta_{m,n}-\delta_{m,n-1}+\delta_{m,n-2}
-\delta_{m,n-3})} \0\\
&&-\,B\,W^{(3)}_{n+2}\,W^{(3)}_{n}\delta_{m,n+1} 
-\,A\,W^{(3)}_{n-2}\,W^{(3)}_{n+1}\delta_{m,n-2}\0\\
W_n^{(1)}\, W^{(3)}_m &=& W^{(3)}_m\, W^{(1)}_n\, q^{\frac{2}{3}(\delta_{m,n+3}+
\delta_{m,n+2}-\delta_{m,n+1}-\delta_{m,n-2})} \label{quantumW}\\
W_n^{(2)}\, W^{(2)}_m &=& W^{(2)}_m\, W^{(2)}_n\, q^{\frac{2}{3}(\delta_{m,n+3}-
\delta_{m,n+1}+\delta_{m,n-1}-\delta_{m,n-3})} \0\\
&&+\,B\,W^{(3)}_{n}\,W^{(1)}_{n+1}\delta_{m,n+1} 
+\,A\,W^{(3)}_{n-1}\,W^{(1)}_{n}\delta_{m,n-1}\0\\ 
W_n^{(2)}\, W^{(3)}_m &=& W^{(3)}_m\, W^{(2)}_n\, q^{\frac{2}{3}(\delta_{m,n+3}+
\delta_{m,n}-\delta_{m,n-1}-\delta_{m,n-2})} \0\\
W_n^{(3)}\, W^{(3)}_m &=& W^{(3)}_m\, W^{(3)}_n\, q^{\frac{2}{3}(\delta_{m,n+2}+
\delta_{m,n+1}-\delta_{m,n-1}-\delta_{m,n-2})} \0
\b
We can recover the classical limit (\ref{Wn}) by expanding in $\hbar$ and
setting $\hbar={1\over 4}$.

We notice that the expressions (\ref{SnWn}) for $S_n,W_n$ are ambiguous in 
the quantum case. One should suitably order them. However, no matter how one
orders them, $W_n^{(3)}$ quantum commutes with them and therefore presumably
constitutes a simple factor of the entire quantum algebra, as in the classical
case.

\section{Discussion}

The results we have obtained above can no doubt be extended to $sl_n$
Toda field theories: in the general case we would obtain a quadratic
algebra with $n$ generators and we would end up with the $n$--KdV hierarchy.
Let us also recall that what we have done in this paper can be done
in the continuum. Compared to the continuum treatment the lattice formalism 
may look awkward but offers some advantages: it provides an automatically 
regularized theory, it allows us to deal with quantum vertex operators (the 
$\sigma_n^i$'s) in a very elegant way -- in general the operator content of 
the quantum theory is better expressed on the lattice -- and it provides a 
more direct connection with integrable hierarchies and matrix models.
In brief continuum and lattice formalisms are usefully complementary
to each other.

Anyhow, both the continuum and the discrete formalism reveal a 
triangular relation among Toda field theories, $n$--KdV
hierarchies of differential equations and matrix models.  
Now let us also recall that the $n$-KdV hierarchies underlie the $A_{n-1}$ 
topological field theory -- in the sense that the correlation functions of 
the latter are determined by the flow equations of the former \cite{DVV}. 
In turn these 
theories are the twisted versions of N=2 supersymmetric Landau--Ginzburg 
models. Therefore we see here {\it a correspondence between Toda field theories 
and N=2 supersymmetric models}. Indications of a contiguity between Toda field
theories and N=2 superconformal models have been found also in \cite{AF}. In 
turn N=2 superconformal field theories brings into the game string vacua. On the
other hand topological field theories, such as those mentioned above, can be 
obtained directly from matrix models, which seem to be natural carriers of 
topological degrees of freedom. We are thus led to a web of 
relationships. Although rather speculative at this stage,
we do not think they are an accidental coincidence. We instead believe that
there is a deep relation among the various facets mentioned above. 
Unfortunately, while each aspect separately has been considerably studied,
a synoptic analysis is still incomplete. For example, it has been shown in 
\cite{W},\cite{K}, that $n$--KdV hierarchies describe topological properties 
of the moduli space of Riemann surfaces. This immediately brings to one's
mind the role of the Liouville equation in the uniformization theory of Riemann
surfaces. A comparison between KdV and the Liouville theory in this 
context is actually tried in \cite{Ma}. But this is the only example, to our 
knowledge. The same can be said about the relations among Toda field theories, 
matrix models and N=2 superconformal field theories. 
But we would not be surprised to 
eventually find out that all these subjects are different facets of the 
same theory. This line of thinking was the original motivation 
of our research.

\subsubsection*{Aknowledgments}
L.P.C. would like to thank ICTP for the kind hospitability and CNPq-
Brazil for financial support. C.P.C. is grateful to FAPESP for financial 
support.

\end{document}